\documentclass[twocolumn,showpacs,amsmath,amssymb,prb]{revtex4}

\usepackage{graphicx}
\usepackage{dcolumn}
\usepackage{bm}

\newcommand{\sr}{$\mbox{SrCu}_{2}\mbox{O}_{3}$}

\begin{document}

\title{Magnetic excitations in $\mbox{SrCu}_{2}\mbox{O}_{3}$: a Raman scattering study}

\author{A. G\"{o}{\ss}ling, U. Kuhlmann, and C. Thomsen}
 \affiliation{Institut f\"{u}r Festk\"{o}rperphysik, Technische Universit\"{a}t Berlin, Hardenbergstr. 36, D-10623 Berlin, Germany}
\author{A. L\"{o}ffert, C. Gross, and W. Assmus}
\affiliation{Physikalisches Institut, J.W.~Goethe-Universit\"{a}t,
Robert-Mayer-Str. 2-4, D-60054 Frankfurt a.~M., Germany}

\date{6 February 2003}

\begin{abstract}
We investigated temperature dependent Raman spectra of the
one-dimensional spin-ladder compound
$\mbox{SrCu}_{2}\mbox{O}_{3}$. At low temperatures a two-magnon
peak is identified at $3160\pm10$ cm$^{-1}$ and its temperature
dependence analyzed in terms of a thermal expansion model. We find
that the two-magnon peak position must include a cyclic ring
exchange of $J_{cycl}/J_{\bot}=0.09-0.25$ with a coupling constant
along the rungs of $J_{\bot} \approx 1215$ cm$^{-1}$ {($1750$ K)}
in order to be consistent with other experiments and theoretical
results.
\end{abstract}


\maketitle

Antiferromagnetic copper oxide spin-ladders have been investigated
intensively from a theoretical and experimental point of view.\cite{Dag96}
Superconducting under high pressure \cite{Ueh96}
they form a bridge between 1D Heisenberg chains and a 2D
Heisenberg square-lattice, which is also a model for high-$T_C$
superconductors. The magnetic ground state of a spin ladder is
surprisingly not a long ranged N\'{e}el state but a short ranged
resonating valence bond state.\cite{And87} The first excited
state is separated from the ground state by a finite energy
$\Delta$. The spin gap was first predicted theoretically
\cite{Dag92,Ric93} and later confirmed experimentally.
\cite{Azu94}
\par
\begin{figure}[b]
\includegraphics[bbllx=55,bblly=446,bburx=505,bbury=778,width=8cm,clip]{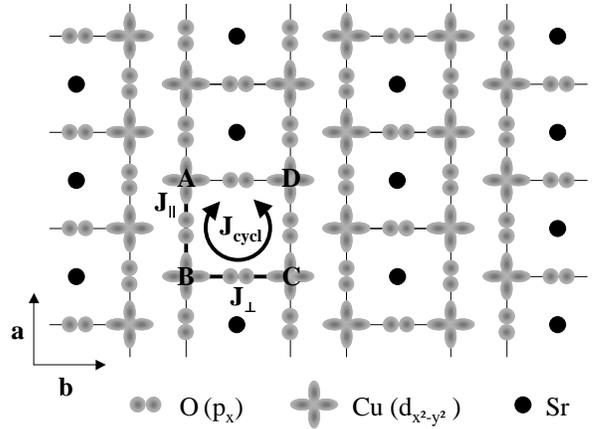}
\caption{\label{fig:super}Schematic view of the two-leg ladder
\sr\ projected on the {$ab$-plane}. The Sr atoms are located in
between the planes containing the $\mbox{Cu}_{2}\mbox{O}_{3}$
atoms. The magnetic coupling constants along the rungs and the
legs are indicated by $J_{\bot}$ and $J_{\|}$ and the cyclic ring
exchange by $J_{cycl}$. The coupling between two Cu {$d$-orbitals}
is caused by superexchange interaction via an O {$p$-orbital.}}
\end{figure}

 Typical realizations of a two-leg spin-ladder are the
compounds \sr\ (Ref.~\onlinecite{Hir91}) and
$\mbox{(Sr,Ca,La)}_{14}\mbox{Cu}_{24}\mbox{O}_{41}$.\cite{Mcc88}
The first compound is a prototype of weakly coupled Cu$_2$O$_3$
{spin-ladders} while the latter consists of a ladder and a Cu$_2$O
edge sharing chain part. In contrast to \sr,
$\mbox{(Sr,Ca,La)}_{14}\mbox{Cu}_{24}\mbox{O}_{41}$ is
intrinsically doped with holes. A schematic view of the compound
\sr\ is shown in Fig.~\ref{fig:super}.
Cu atoms, represented by {$d$-orbitals}, are antiferromagnetically
coupled via an intermediate oxygen {$p$-orbital} by superexchange
\cite{And50}. The Sr atoms are located in between the planes
containing the $\mbox{Cu}_{2}\mbox{O}_{3}$ atoms. The coupling
constants along the legs and the rungs are denoted with $J_{\|}$
and $J_{\bot}$. The interladder coupling is negligible to first
approximation because the superexchange via a {Cu-O-Cu} path with
a 90$^\circ$ bond angle has a smaller orbital overlap than with a
bond angle of 180$^\circ$.\cite{Gra99} Thus, a ladder can be
considered an isolated quasi one-dimensional object with a
Heisenberg Hamiltonian $H=J_{\bot}\sum_{\text{rung}}\textbf{S}_i
\cdot \textbf{S}_j+J_{\|}\sum_{\text{leg}}\textbf{S}_i \cdot
\textbf{S}_j$. In the literature coupling ratios $J_{\|}/J_{\bot}
\sim 1.7-2.0$ have been reported \cite{Ecc98,Ogi00} taking the
Hamiltonian mentioned above into account for analyzing the data.
On the other hand with almost identical Cu-Cu distances in leg and
rung direction \cite{Hir91,Kaz97,Mcc88} one would expect an
isotropic ratio $J_{\|}/J_{\bot} \sim 1$. This picture was also
confirmed by an analysis of optical conductivity data.\cite{Win01}
Recently the inclusion of a cyclic ring exchange
$J_{cycl}$ was suggested in order to resolve the discrepancy
between the geometrical considerations and $J_{\|}/J_{\bot} \sim
1.7-2.0$.\cite{Bre99,Mat00,Nun02} This ring exchange can be
understood as a superposition of clockwise and counter clockwise
permutations of four spins around a plaquette (positions ABCD in
Fig.~\ref{fig:super}). A term
$H_{cycl}=J_{cycl}\sum_{i}K^i_{ABCD}$ with
$K_{ABCD}=(\textbf{S}_A\textbf{S}_B)(\textbf{S}_C\textbf{S}_D)+
(\textbf{S}_A\textbf{S}_D)(\textbf{S}_B\textbf{S}_C)-(\textbf{S}_A
\textbf{S}_C)(\textbf{S}_B\textbf{S}_D)$ has to be added to the
conventional Heisenberg Hamiltonian $H$.\cite{Sch02} In order to
achieve the isotropic limit of $J_{\|}/J_{\bot} \sim 1$ a ring
exchange of $J_{cycl}/J_{\bot} \sim 0.18-0.30$ was introduced.\cite{Mat00,Nun02} \par In this paper we studied the Raman
spectra of the two leg $S=\frac{1}{2}$ spin-ladder compound \sr.
In addition to phonons we investigated the two-magnon peak in this
compound at temperatures between 25 K and 300 K. We find that the
inclusion of a ring exchange is necessary for the understanding of
the magnetic properties of \sr.
\par
\begin{figure}
\includegraphics[width=6.5cm,clip]{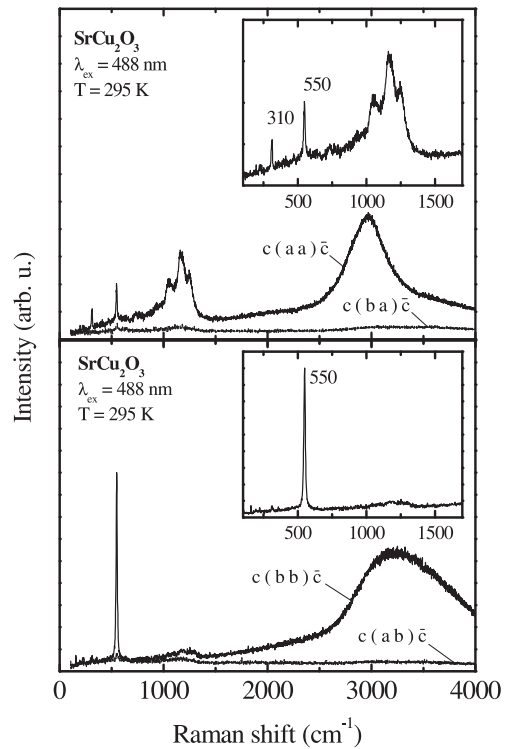}
\caption{\label{fig:sr2}Raman spectra in parallel and crossed
polarization measured at $T$=295 K. The insets show the low energy
parts of the spectra measured in parallel polarization.}
\end{figure}

\begin{figure}
\includegraphics[width=6.5cm,clip]{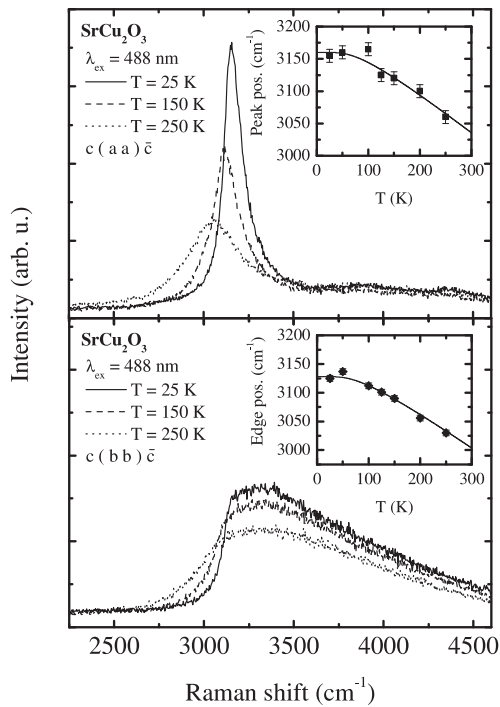}
\caption{\label{fig:sr}Raman spectra measured with the same
polarization as in Fig.~\ref{fig:sr2} with $\lambda_{ex}=488$ nm
at 25 K (solid), 150 K (dashed) and 250 K (dotted). Insets: peak
positions (top) and edge position (bottom) of the two-magnon peak
as a function of temperature.}
\end{figure}

 Polycrystalline \sr\ was grown under high pressure as
described by L\"{o}ffert \textit{et al.}\cite{Loe02} The
crystallographic structure was verified by x-ray diffraction. In
addition to \sr\ small amounts of
$\mbox{Sr}_2\mbox{Cu}_3\mbox{O}_5$, $\mbox{Cu}\mbox{O}$ and
$\mbox{Cu}_{2}\mbox{O}$ were detected. Measurements were performed
using a LabRam spectrometer (Dilor) including a grating with 600
and 1800 grooves/mm in backscattering geometry and a multichannel
CCD detector. An Ar$^+$ laser spot (488 nm) was focused by a 80
$\times$ microscope objective on small individual crystallites on
the sample surface with a diameter of 1-2 $\mu$m. The spectra in
Fig.~\ref{fig:sr2} were measured at room temperature. The
crystallite was chosen by using a polarization microscope
:\cite{Sug99a} The sample surface was illuminated normally with
linear polarized white light. The reflected light was detected
through an analyzer crossed to the polarizer with a CCD camera.
Being an anisotropic material the reflected light is in general
elliptically polarized. As for the ladder compound
$\mbox{LaCuO}_{2.5}$,\cite{Sug99a} we assume the direction along
the ladder ($a$-axis) to be the one with the largest anisotropy in
\sr, whereas the $b$ and $c$ axes are approximately optically
isotropic. Thus, there should be a difference in brightness along
and perpendicular to the $a$-axis when rotating the sample. If the
incident wave vector $\textbf{k}_{i}$ is parallel to the $a$-axis
the image stays dark while rotating the sample. For
$\textbf{k}_{i} \perp a$ the crystallite appears
dark and bright for four times when turning the sample $360^{\circ}$.
The spectra in Fig.~\ref{fig:sr2} were measured on a crystallite
for $\textbf{k}_{i} \perp a$. For our measurements down to 25 K
(spectra in Fig.~\ref{fig:sr}) the sample was mounted on the cold
finger of an Oxford microcryostat.

The Raman spectra of \sr\ can be divided into a low (100-1700
cm$^{-1}$) and a high-energy (2000-4600 cm$^{-1}$) region as shown
in Figs.~\ref{fig:sr2} and ~\ref{fig:sr}. We verified the chemical
composition of our polycrystalline sample by an analysis of the
phonons in the low energy part of the spectra. We start with a
factor group analysis (FGA) of the vibrational modes according to
Rousseau \textit{et al.}\cite{Rou81} The unit cell of \sr\
consists of two formula units. Structure analysis
\cite{Hir91,Joh00} attributes \sr\ to the space group $Cmmm$
($D_{2h}^{19}$). The FGA results in
\begin{equation}
\Gamma_{\mbox{SrCu}_{2}\mbox{O}_{3}}=2A_{g}+4B_{1u}+2B_{1g}+4B_{2u}+4B_{3u}+2B_{3g}
\end{equation}
Having a center of inversion only even modes show Raman activity,
resulting in six Raman active phonons. We expect the two $A_g$
modes in the spectra measured in parallel polarization while the
four even $B$ modes should be observed in crossed polarization.
\par Figure~\ref{fig:sr2} shows four spectra measured at {$T$=295
K} two in parallel and two in crossed polarization. The axes were
identified  using the described analysis with polarized light;
comparing Fig.~\ref{fig:sr2} to the spectra of the very similar
ladder compound $\mbox{Sr}_{14}\mbox{Cu}_{24}\mbox{O}_{41}$,\cite{Pop00,Sug99,Goz01,Ogi00} we identify the $a$ (leg) and $b$
(rung) axes as indicated in the figure.\cite{exp}
\par In $(bb)$ polarization only a single phonon at 550
cm$^{-1}$ is observed, while in $(aa)$ two phonons at 310, 550
cm$^{-1}$, and in addition several broad peaks around 1150
cm$^{-1}$ are visible. We assign these two phonons as the two
allowed $A_g$ modes. In
$\mbox{Sr}_{14}\mbox{Cu}_{24}\mbox{O}_{41}$ modes at 316 cm$^{-1}$
and 549 cm$^{-1}$ were observed.\cite{Pop00} The latter is
believed to be the breathing mode of the O-ladder atoms located on
the ladder legs \cite{Pop00}, which is in accordance with our
assignment. The structure at about 1150 cm$^{-1}$ was also
observed in $\mbox{Sr}_{14}\mbox{Cu}_{24}\mbox{O}_{41}$, its
origin attributed to two-phonon processes.\cite{Abr97,Pop00}
Knowing only few phonons at the $\Gamma$ point the details of the
broad structure around 1150 cm$^{-1}$ in \sr\ remain unclear. In
summary, we identified the two allowed  $A_g$ ladder phonons at
310 and 550 cm$^{-1}$ in \sr\ at room temperature, the $B_{1g}$
modes were too weak to be observed, and the $B_{3g}$ not allowed
for our geometry.
\par In Fig.~\ref{fig:sr2} in addition to the phonons a peak
around 3000 cm$^{-1}$ was observed in $(aa)$ and $(bb)$
polarizations while it was absent in the crossed polarizations
$(ab)$ and $(ba)$. It can be shown from the two-magnon Raman
Hamiltonian \cite{Fle68} that for a two-leg Heisenberg ladder the
two-magnon peak is forbidden in all crossed polarizations. The
observed spectra follow the selection rules of a two-magnon peak.

In Fig.~\ref{fig:sr} the Raman spectra of the high-energy region
are presented for different temperatures. The spectra were
measured on another crystallite between 25 K and 300 K. Comparing
them to the aligned spectra in Fig.~\ref{fig:sr2} we assign the
upper spectrum as $(aa)$ and the lower one as $(bb)$ polarized. In
$(aa)$ polarization a sharp peak is visible which broadens and
shifts to lower energies with increasing temperature. In $(bb)$
polarization a much broader peak is measured; it remains broad
also down to the lowest temperatures. We identify these features
to be the two-magnon peaks which were also observed in
$\mbox{Sr}_{14}\mbox{Cu}_{24}\mbox{O}_{41}$ at a lower energy
(maximum for $T$=8-20 K at 2900-3000 cm$^{-1}$).\cite{Pop00,Sug99,Goz01,Ogi00} The energy of the peak maxima as a
function of temperature is shown in the upper inset of
Fig.~\ref{fig:sr}. In $(bb)$ polarization the left edge position
is plotted as a function of temperature in the lower insets of
Fig.~\ref{fig:sr} due to difficulties in determining the precise
peak maxima. The two-magnon peak shifts for $T>$75 K almost
linearly with temperature and saturates at low temperatures at
3160 cm$^{-1}$ (peak position) and 3130 cm$^{-1}$ (edge position).
\par Using a simple model of thermal expansion we are able to
understand the temperature dependence of these peaks. The starting
point is a system containing two copper {$d$-orbitals} and one
intermediate oxygen {$p$-orbital} with a $180^{\circ}$ bond angle
as found in flat ladder compounds, hence the model is not only
applicable to \sr\ but also, e.g., to
$\mbox{Sr}_{14}\mbox{Cu}_{24}\mbox{O}_{41}$. The two copper atoms
are coupled by superexchange with the coupling constant $J$. We
assume that the two-magnon energy scales linearly with $J$. The
average distance $d$ between two copper atoms is a function of
temperature. Due to an anharmonic interatomic potential the
average distance increases in first order as $d=d_{0}+{A
\omega}/{(\exp({\omega/T})-1)}$, where $d_{0}$ is the distance at
{$T$=0 K}, $A$ is a constant, representing the strength of the
expansion, and $\omega$ is an averaged phonon energy (Einstein
model). Within the three-band Hubbard-model $J$ equals
${4t_{pd}^4}/{\varepsilon_{pd}^2}\: ({U_d}^{-1}+{\varepsilon_{pd}^{-1}})$ with $t_{pd}$ being the
overlap integral between Cu-O sites, $U$ the Coulomb repulsion,
and $\varepsilon_{pd}$ the charge transfer energy.\cite{Mue02}
The overlap integral $t_{pd} \propto d^{-4}$ and the Coulomb
repulsion $U \propto d^{-1}$ are a functions of $d$ as shown by
Harrison.\cite[p. 643]{Har99} Using the parameters $C_{1,2}$, $J$
can be written as
\begin{equation}
J = C_1 d^{-15} + C_2 d^{-16} \label{eq:power}
\end{equation}
The temperature dependence of the two-magnon energy, which is
proportional to $J$, arises from a Taylor expansion around $d_0$.
\begin{equation}
{\label{equ:T}}
E_{2-mag}={E_{0}}-\frac{B\omega}{\exp({\omega/T})-1}
\end{equation}
$E_0$ is the two-magnon energy at $T$=0 K, $B$ is a dimensionless
constant proportional to $A$, and $\omega$ is the averaged phonon
frequency. Other power-law exponents in Eq.~\ref{eq:power} have
been used in similar systems in the literature (e.g., Kawada
\textit{et al.}\cite{Kaw98}); the further analysis does not
depend much on this choice as long $B$ is simply a parameter. The
temperature dependence of the peak/edge maxima of both spectra in
the insets of Fig.~\ref{fig:sr} were fitted simultaneously using
Eq.~(\ref{equ:T}) with the parameters $B=0.94 \pm 0.05$,
$\omega=160 \pm 10$ cm$^{-1}$, ${E_{0}}(aa)=3160 \pm 10$ cm$^{-1}$
and ${E_{0}}(bb)=3130 \pm 10$ cm$^{-1}$. The fit shows excellent
agreement with the experimental data. The averaged phonon
frequency $\omega$ is comparable to those of high-$T_C$ cuprate
materials.\cite{Kno90} The different values for $E_0$ in the
$(aa)$ and $(bb)$ spectra in Fig.~\ref{fig:sr} are caused by the
difference between peak edge and peak maximum. For the further
analysis we take only the energy ${E_{0}}(aa)$=3160 cm$^{-1}$ into
account. The observed temperature dependence is in accordance with
the assumed magnetic origin of the peak. The common picture of
superexchange is confirmed by our measurements.
\par We now show that in order to compare the energy $E_0$ with
existing experiments and theoretical results in a consistent way
the cyclic ring exchange must be included. Note, that it is not
possible to determine the magnetic constants and the spin gap
directly from the two-magnon peak position. In order to obtain
these values from our spectra we have to make several assumptions
based on calculations: (i) \textit{Ab initio} calculations
\cite{Gra99} yielded $J_{\|}/J_{\bot}=1.1$ for the coupling
constant ratio. This ratio is also close to the isotropic limit,
which can be deduced from geometrical considerations. (ii) We
include a ring exchange of $J_{cycl}/J_{\bot}=0.09 - 0.25$. Cyclic
ring exchange values on the same order
($J_{cycl}/J_{\bot}=0.18-0.30$) have also been introduced for the
comparable two-leg spin ladder compound
$\mbox{(La,Ca)}_{14}\mbox{Cu}_{24}\mbox{O}_{41}$
(Refs.~\onlinecite{Mat00,Nun02}). (iii) Schmidt \textit{et al.} have shown
recently in a theoretical Raman study for a Heisenberg two-leg
ladder with $J_{\|}/J_{\bot}=1.2$ and $J_{cycl}/J_{\bot}=0.2$ that
the two-magnon peak maximum should be located at $E_0 \approx 2.6
\cdot J_{\bot}$.\cite{Sch02} Because the ratio
$J_{\|}/J_{\bot}=1.2$ is slightly larger than the one stated above
for \sr\ we take $E_0 / J_{\bot} = 2.6 \pm 0.1$ for a further
analysis. (iv) Calculating the spin gap we use the expression
$\Delta \approx 0.48 \cdot J_{\bot}-1.08 \cdot J_{cycl}$ which has
been derived by Nunner \textit{et al.}\cite{Nun02} for a two-leg
Heisenberg ladder with ring exchange having adjusted for the
different definitions of $J_{cycl}$. The expression for $\Delta$
is valid approximately in the range $J_{\|}/J_{\bot}=1.0-1.3$. As
a result we obtain from our experiment $J_{\bot} = 1170-1260$
cm$^{-1}$ ($1680-1810$ K) for the coupling constant. Taking
$J_{\bot}=1215$ cm$^{-1}$ results in $\Delta =260-470$ cm$^{-1}$
($380-680$ K) for the spin gap. The latter value is in good
agreement with spin gap values measured with susceptibility (420
K),\cite{Azu94} NMR (680 K)\cite{Azu94,Ish94,Ish96} and neutron
scattering (380 K) on $\mbox{Sr}_{14}\mbox{Cu}_{24}\mbox{O}_{41}$.
\cite{Ecc98} In contrast, excluding the spin exchange coupling
would yield a spin gap value of approximately $840$ K, which is
out of scale by a factor of $1.2-2.2$ when comparing to the
experiments mentioned. Our experiment thus supports the presence
of a four spin ring exchange on the order of
$J_{cycl}/J_{\bot}=0.09-0.25$. The lower limit of
$J_{cycl}/J_{\bot}$ so obtained corresponds to $\Delta$ from NMR,
the upper one to $\Delta$ from neutron scattering.

In conclusion we found the two allowed $A_g$ phonons at 310
cm$^{-1}$ and 550 cm$^{-1}$ in \sr\ and studied the magnetic
excitations in this compound. A peak with an energy of 3160$\pm
10$ cm$^{-1}$ at low temperatures starts to shift to lower
energies with increasing temperature. We identified this peak as
the two-magnon peak and confirmed this by its temperature
dependence. We derived a simple expression which describes
excellently the peak energy as a function of temperature. For low
temperatures we are able to explain the two-magnon peak energy
$E_0$=3160 cm$^{-1}$ in accordance with existing theoretical and
experimental results by including a cyclic ring exchange using the
values: $J_{\|}/J_{\bot}=1.1$, $J_{cycl}/J_{\bot}=0.09-0.25$,
$J_{\bot} \approx 1215$ cm$^{-1}$ (1750 K) and $\Delta$=260-470
cm$^{-1}$ ($380-680$ K).

We thank K.~P. Schmidt and G.~S. Uhrig for helpful and valuable
discussions. This work was supported by the Deutsche
Forschungsgemeinschaft, SPP 1073.

%\*****************************************************************************************


\begin{thebibliography}{30}
\bibitem{exp} Note that the crystallographic directions of the rungs and legs are
different in other ladder compounds.
\end{thebibliography}
\end{document}